\documentclass[twocolumn,prd,floatfix,showpacs]{revtex4}
\usepackage{graphicx}
\usepackage{epsfig}
\usepackage{psfrag}

\begin{document}


\title{Can Planck-scale physics be seen in the cosmic microwave background  ?}
\author{{\O}ystein Elgar{\o}y}
\affiliation{NORDITA, Blegdamsvej 17, DK-2100 Copenhagen, Denmark}
\email{oelgaroy@nordita.dk}
\author{Steen Hannestad}
\affiliation{Department of Physics, University of Southern Denmark, 
Campusvej 55, DK-5230 Odense M, Denmark \\
and
\\
NORDITA, Blegdamsvej 17, DK-2100 Copenhagen, Denmark}
\email{hannestad@fysik.sdu.dk}
\date{{\today}}

\begin{abstract}
We investigate the potential of observations of anisotropies in 
the cosmic microwave background (CMB) and large-scale structure in the 
Universe to detect possible modifications of standard inflationary
models by physics beyond the Planck scale.  A generic model of the 
primordial density fluctuations is investigated, and we derive
constraints on its parameters from current data.  We conclude that 
the currently available data do not put very stringent constraints 
on this model.  Furthermore, we use simulated power spectra from 
the Sloan Digital Sky Survey (SDSS) and the Planck satellite to show 
that it is unlikely that a trans-Planckian signature of this type 
can be detected in CMB and large-scale structure data.   

\end{abstract}
\maketitle


\section{Introduction}
Quantum gravitational effects are expected to modify the spectrum of 
primordial density fluctuations produced during the inflationary phase
in the very early Universe.  Since the primordial fluctuations are the 
seeds for the anisotropies in the cosmic microwave background (CMB) 
and for the large-scale structures we observe in the Universe,   
cosmological observations have the potential to shed light on  
on Planck-scale physics.  
However, although enormous 
progress has been made in observational cosmology in the past few years, 
it is a highly non-trivial task to separate the primordial 
density fluctuations from the present-day, processed power spectra 
which we can measure.  In this paper we show that the  
currently available data do not impose any stringent constraints on 
trans-Planckian modulations of the primordial power spectrum. 
Furthermore, we investigate whether  future, more precise measurements 
will improve the prospects for detecting a 
signature for quantum gravity.
Based on suggestions for the form and size of trans-Planckian 
effects on the primordial power spectrum (see
e.g. \cite{brandenberger,kempf,kaloper,easther,danielsson1,danielsson2,alberghi}), 
we consider a  model 
where a power-law spectrum is modulated by an oscillating function of 
the comoving
wavenumber $k$ \cite{danielsson1,danielsson2,alberghi}.   
By performing a full likelihood analysis of simulated data sets, we 
are led to the conclusion that it will be extremely difficult to detect 
trans-Planckian effects in CMB and large-scale structure data. 

The structure of this paper is as follows.  In section II we give 
a brief summary of the work on trans-Planckian effects on the 
primordial power spectrum and introduce our model.  
Section III contains a discussion of the cosmological data, with 
an emphasis on the unavoidable smearing of power by the window
functions.  Section IV describes our likelihood analysis and the 
results, 

\section{Trans-Planckian physics and the primordial power spectrum}

The standard calculation of the perturbations produced during the 
inflationary phase is based on flat-space quantum field theory, and
the initial conditions for the inflaton $\phi$ are imposed in the 
infinite past.  Since the Universe is expanding, this means that 
any given Fourier mode $k$ was infinitesimally small at that time.  
Since our expectation is that `new physics' becomes important on 
length scales below the Planck length, the standard calculation is 
probably too naive, and we should expect corrections from quantum
gravitational effects.  We do not 
know what the correct fundamental theory on scales comparable to and 
smaller than the 
Planck length is, and hence it is impossible to predict these corrections 
with a high degree of certainty.  
One can only devise plausible scenarios based on informed 
guesses as to what the significant features of a theory of quantum 
gravity should be like.  Many proposals have been
made with a varying degree of optimism regarding possible observable 
signatures \cite{brandenberger,kempf,kaloper,easther,danielsson1,danielsson2,alberghi}.  
An important point is 
whether the leading-order correction to the primordial power spectrum 
is linear in $H/\Lambda$, where $H$ is the Hubble constant during
inflation 
and $\Lambda$ is the energy scale where new physics enters.  
If the correction is of higher order than linear,
it will certainly be undetectable, so we will assume that it is linear.  
With one specific choice of vacuum, 
it was shown in \cite{danielsson1,danielsson2} that the primordial 
power spectrum modified by trans-Planckian effects is given by 
\begin{equation}
P(k;\epsilon,\xi) =
P_0(k)\left\{1-\xi\left(\frac{k}{k_n}\right)^{-\epsilon}
\sin\left[\frac{2}{\xi}\left(\frac{k}{k_n}\right)^\epsilon\right]
\right\},
\label{eq:transplanckpk1}
\end{equation}
where $\epsilon$ is the standard inflationary slow-roll parameter 
related to the inflaton potential $V(\phi)$ by 
$\epsilon=(16\pi)^{-1}M_{\rm P}^2(V'/V)^2$, $M_{\rm P} 
= 1/\sqrt{8\pi G}$ is the reduced Planck mass, 
$\xi \sim 4\times 10^{-4} 
\sqrt{\epsilon}/\gamma$, $\gamma = \Lambda / M_{\rm P}$ parametrizes 
the scale where trans-Planckian effects enter, and $P_0(k)$ is the 
primordial power spectrum predicted by standard inflationary theory, 
typically given by the scale-free form $P_0(k)\propto k^{n_s-1}$, where 
$n_s$ is the scalar spectral index.  Finally, $k_n$ corresponds to the
largest scales measurable in the CMB.  The effect is seen to be a
modulation of $P_0(k)$, and the natural question to ask is whether
this is detectable.  

Based on simulated data for the Planck satellite \cite{planck}, it was 
concluded in \cite{danielsson2} that the modulation
should be detectable in the  CMB data from this upcoming mission.  
However, this conclusion was based on a Fisher matrix 
analysis, not on a full likelihood analysis.  
In fact, as we will show, the full likelihood analysis reveals 
that the issue is more complicated.  The likelihood oscillates in 
parameter space, and one can be misled if one looks only at 
its local behaviour.

\section{Observational data}

\subsection{Cosmic microwave background}

The CMB temperature
fluctuations are conveniently described in terms of the
spherical harmonics power spectrum
\begin{equation}
C_l \equiv \langle |a_{lm}|^2 \rangle,
\end{equation}
where
\begin{equation}
\frac{\Delta T}{T} (\theta,\phi) = \sum_{lm} a_{lm}Y_{lm}(\theta,\phi).
\end{equation}
Since Thomson scattering polarizes light there are additional power spectra
coming from the polarization anisotropies. The polarization can be
divided into a curl-free $(E)$ and a curl $(B)$ component, yielding
four independent power spectra: $C_{T,l}, C_{E,l}, C_{B,l}$ and 
the temperature $E$-polarization cross-correlation $C_{TE,l}$.

The WMAP experiment have reported data  on $C_{T,l}$ and $C_{TE,l}$,
as described in Ref.~\cite{map1,map2,map3,map4,map5}

We have performed the likelihood analysis using the prescription
given by the WMAP collaboration which includes the correlation
between different $C_l$'s \cite{map1,map2,map3,map4,map5}. Foreground contamination has
already been subtracted from their published data.

In parts of the data analysis we also add other CMB data from
the compilation by Wang {\it et al.} \cite{wang3}
which includes data at high $l$.
Altogether this data set has 28 data points.

\subsection{Large scale structure}

The 2dF Galaxy Redshift Survey (2dFGRS) \cite{colless} has measured 
the redshifts 
of more than 230 000 galaxies with a median redshift of 
$z_{\rm m} \approx 0.11$.  
An initial estimate of the convolved, redshift-space power spectrum of the 
2dFGRS has been determined \cite{percival} for a sample of 160 000 redshifts. 
On scales $0.02 < k < 0.15h\;{\rm Mpc}^{-1}$ the data are robust and the 
shape of the power spectrum is not affected by redshift-space or nonlinear 
effects, though the amplitude is increased by redshift-space distortions.  
A potential complication is the fact that the galaxy power spectrum 
may be biased with respect to the matter power spectrum, i.e. light does not 
trace mass exactly at all scales.  This is often parametrised by introducing 
a bias factor 
\begin{equation}
b^2(k)\equiv \frac{P_{\rm g}(k)}{P_{\rm m}(k)}, 
\label{eq:biasdef}
\end{equation}
where $P_{\rm g}(k)$ is the power spectrum of the galaxies, and $P_{\rm m}(k)$ 
is the matter power spectrum.    
However, we restrict our analysis of the 2dFGRS power spectrum to scales 
$k < 0.15 \;{\rm h}\,{\rm Mpc}^{-1}$ 
where the power spectrum is well described by linear theory.   
On these scales, two different analyses have demonstrated  
that the 2dFGRS power spectrum is consistent with linear, 
scale-independent bias \cite{lahav,verde}.   
Thus, the shape of the galaxy power spectrum 
can be used straightforwardly to constrain the shape of the matter power 
spectrum.  
However, when looking for modulations or other features in the 
primordial power spectrum 
using the 2dFGRS, one should bear in mind that what is measured is the 
convolution of the true galaxy power spectrum with the 2dFGRS window function 
 $W$ \cite{percival}, 
\begin{equation}
P_{\rm conv}({\bf k}) \propto \int P_{\rm g}({\bf k}-{\bf q})
|W_k({\bf q})|^2 d^3q,  
\label{eq:convolve1}
\end{equation}
and it was found in \cite{elgaroy}  
that this convolution washes out any features in 
the primordial power spectrum for $k < 0.1h\;{\rm Mpc}^{-1}$. 
However, combining the 2dFGRS power spectrum with CMB data breaks parameter 
degeneracies that are present if each dataset is analysed separately, and 
therefore a combination of large-scale structure and CMB data gives 
tighter constraints on the primordial power spectrum than the CMB alone. 

\subsection{Likelihood analysis}

For calculating the theoretical CMB and matter power spectra we use
the publicly available CMBFAST package \cite{CMBFAST}.
As the set of cosmological parameters we choose
$\Omega_m$, the matter density, 
$\Omega_b$, the baryon density, $H_0$, the
Hubble parameter, $\tau$, the optical depth to reionization,
$Q$, the normalization of the CMB power spectrum, $b$, the 
bias parameter, and finally $\epsilon$ and $\xi$. 
We restrict the analysis to geometrically flat models
$\Omega_m + \Omega_\Lambda = 1$.

In principle 
one might include even more parameters in the analysis, such as
$r$, the tensor to scalar ratio of primordial fluctuations. However, $r$
is most likely so close to zero that only future high precision 
experiments may be able to measure it. The same is true for other 
additional parameters. Small deviations from slow-roll during inflation
can show up as a logarithmic correction to the simple power-law
spectrum predicted by slow-roll.
\cite{Hannestad:2000tj,Hannestad:2001nu,griffiths}
or additional relativistic energy
density 
\cite{Jungman:1995bz,Lesgourgues:2000eq,Hannestad:2000hc,Esposito:2000sv,
Kneller:2001cd,Hannestad:2001hn,Hansen:2001hi,Bowen:2001in,Dolgov:2002ab}
could be present. However, there is no evidence of any such effect in
the present data and therefore we restrict the analysis to the
``minimal'' standard cosmological model.

In this full numerical likelihood analysis we use the free parameters
discussed above with certain priors determined from cosmological
observations other than CMB and LSS.
In flat models the matter density is restricted by observations
of Type Ia supernovae to be $\Omega_m = 0.28 \pm 0.14$ 
\cite{Perlmutter:1998np}.
The current estimated range for $\Omega_b h^2$ from BBN is 
$\Omega_b h^2 = 0.020 \pm 0.002$ \cite{Burles:2000zk}, 
and finally the HST Hubble key
project found the value of $H_0$ to be $72 \pm 8 \,\, {\rm km} \, 
{\rm s}^{-1} \, {\rm Mpc}^{-1}$ \cite{freedman}.
The marginalisation over parameters
other than $\epsilon$ and $\xi$ 
was performed using a simulated annealing
procedure \cite{Hannestad:wx}.

\section{Results} 

\begin{figure}[t]
\vspace*{-0.0cm}
\begin{center}
\epsfysize=7truecm\epsfbox{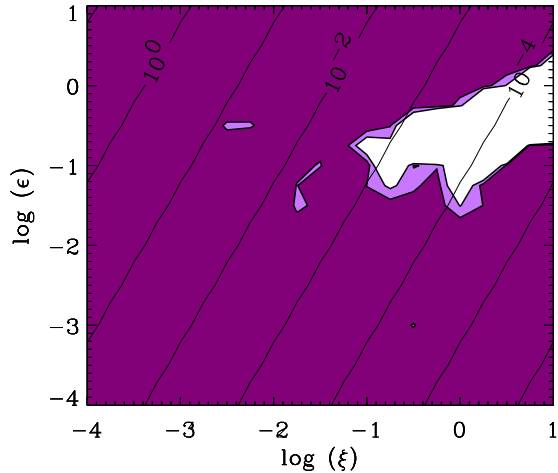}
\end{center}
\caption{68\% and 95\% confidence exclusion plot of the parameters
$\epsilon$ and $\xi$ for WMAP data only. The straight lines are
isocontours for the parameter $\gamma$.} 
\label{fig:wmaponly}
\end{figure}

\begin{figure}[t]
\vspace*{-0.0cm}
\begin{center}
\epsfysize=7truecm\epsfbox{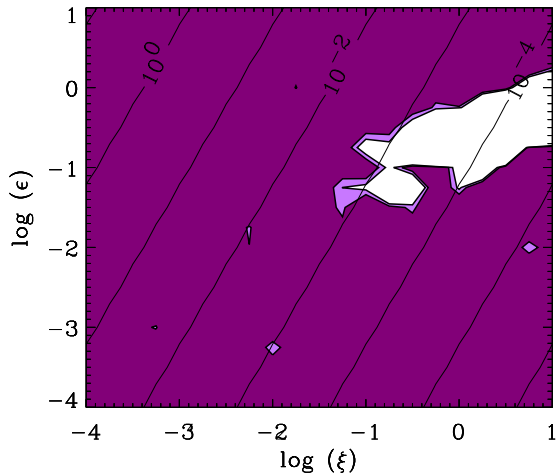}
\end{center}
\caption{68\% and 95\% confidence exclusion plot of the parameters
$\epsilon$ and $\xi$ for the WMAP data set combined with the pre-WMAP
data compilation from Wang et al. The straight lines are
isocontours for the parameter $\gamma$.} 
\label{fig:wmapwang}
\end{figure}

\begin{figure}[t]
\vspace*{-0.0cm}
\begin{center}
\epsfysize=7truecm\epsfbox{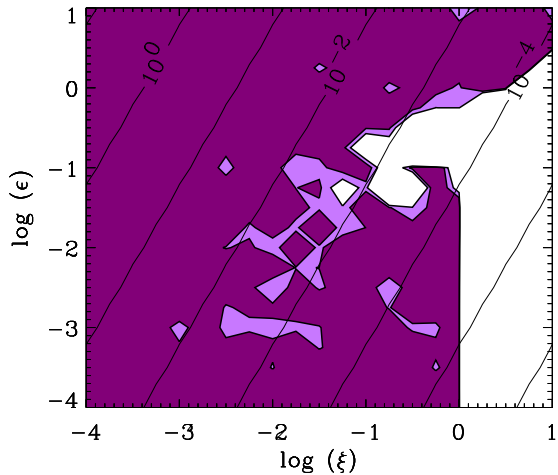}
\end{center}
\caption{68\% and 95\% confidence exclusion plot of the parameters
$\epsilon$ and $\xi$ with the previous CMB data plus additional data
from the 2dF galaxy survey. The straight lines are
isocontours for the parameter $\gamma$.} 
\label{fig:alldata}
\end{figure}

In this section we present the results of the likelihood analysis of our 
selection of current cosmological data sets.  First of all, in figure 
\ref{fig:wmaponly} we show 68 \% and 95 \% exclusion limits in the 
$\xi$--$\epsilon$ parameter space (all other parameters have been 
marginalised over) for the WMAP TT and TE power spectra only.   
Figure \ref{fig:wmapwang} shows how the contraints 
change when the compilation of pre-WMAP CMB data from \cite{wang3}, 
and in figure \ref{fig:alldata} the 2dFGRS power spectrum data have 
been included as well.

From these figures it is clear that there is no preference in the
data for an oscillating power spectrum, at any significant level.
The $\epsilon = \xi = 0$ point is always within the 1$\sigma$ allowed
range.

For the CMB-only data the current excluded range is for $\xi \gtrsim 0.1$
and $0.1 \gtrsim \epsilon \gtrsim 1$. 
$\xi$ is a measure of the amplitude of oscillations, and the bound on
this parameter comes from the fact that small amplitude oscillations
are not visible, given the precision of current experiments.
Since $\epsilon$ roughly
determines the period of the power spectrum oscillations it is 
clear why only a very limited range is excluded. For small values of 
$\epsilon$ the power spectrum does not go through a whole oscillation
within the visible range of $k$-values. On the other hand, for very large 
$\epsilon$ the oscillation period becomes small compared to the
width of the power spectrum window function.

In the case where LSS data is included the conclusion is very similar,
except for the presence of an excluded region beyond $\xi = 1$ at low
$\epsilon$. The reason for this simply is that when $\xi > 1$ the
power spectrum can take negative values for some $k$. Since the 
amplitude fitting of the spectrum is done my multiplying with a positive,
real number this means that no good fit can possibly be obtained when
starting from an underlying power spectrum with negative values.
Of course it also reflects the transition to the unphysical region
of parameter space since the power spectrum should be a positive
definite quantity.



\section{Simulated, high--precision data sets}

\subsection{Simulated Planck data}

We have simulated a data set from the future Planck mission using
the following very simple prescription:
We assume it to be cosmic variance (as opposed to foreground) limited
up to some maximum $l$-value, $l_{\rm max}$, which we take to be
2000.  
For the sake of simplicity we shall work only with the temperature
power spectrum, $C_{T,l}$, in the present analysis.
In fact the Planck detectors will be able to measure polarization
as well as temperature anisotropies, but our simplification of using
only $C_{T,l}$ will not have a significant qualitative impact on our
conclusions. For a cosmic variance limited full sky experiment the 
uncertainty in the measurement of a given $C_l$ is simply
\begin{equation}
\frac{\sigma (C_l)}{C_l} = \sqrt{\frac{2}{2l+1}}.
\end{equation}
It should be noted that taking the data to be cosmic variance limited
corresponds to the best possible case. In reality it is likely 
that foreground effects will be significant, especially at high
$l$ (see Ref.~\cite{foregrounds}). Therefore, our estimate of the
precision with which the oscillation parameters can be measured
is probably on the optimistic side.

\subsection{Simulated SDSS data}

The Sloan Digital Sky Survey (SDSS) \cite{sdss} aims at measuring the
redshift of approximately 1 million galaxies, and from this the 
galaxy power spectrum can be obtained with unprecedented accuracy.  
We will consider how this improves our ability to 
constrain trans-Planckian physics, assuming  
that the bias of galaxies with respect to dark matter is simple and 
scale-independent.

For a fixed survey strategy, the survey volume and the number density
of galaxies in the redshift sample set a lower bound on the
uncertainty in the estimated power spectrum.  In the limit of a 
perfectly spherical volume-limited sample, the uncertainty in the 
estimated power per mode is roughly 
\begin{equation}
\frac{\delta \overline{P}(k)}{\overline{P(k)}} \approx \sqrt{2\frac{V_c}{V_k}}
\left[1+\frac{S(k)}{P(k)}\right], 
\label{eq:sloanerr}
\end{equation}
\cite{fkp} where $P(k)$ and $\overline{P}(k)$ are the true and
estimated power spectra, $S(k) = 1/\overline{n}$ 
is the shot noise
power for a mean galaxy density $\overline{n}$
(we will use $\overline{n}=2\times 10^{-3}\;h^3 \,{\rm Mpc}^{-3}$ ), 
$V_c=(2\pi)^3/V_S$ is 
the coherence volume in the Fourier domain for a survey with volume
$V_S$, 
and it is assumed that we average the power estimates over a shell in 
Fourier space with volume $V_k \approx 4\pi k^2 \Delta k$.  That is, 
we average the power over all angles, and over bins with $\Delta k >
2\pi/R$, 
where $R$ is the survey depth, which we will take to be $500\;h\,{\rm
  Mpc}^{-1}$.    

In constructing a mock window function we follow \cite{fkp}, so  
the window function in Fourier space is the Fourier transform of the 
product of the survey mask with the redshift selection function 
$\varphi(r)$.  Since we are anyway interested in constructing an 
optimistic estimate of the window function, we assume that the 
survey geometry is simple and has full coverage of $2\pi$ steradians,   
\cite{cole98} and adopt a magnitude limit $B_J < 18.9$.   

\begin{figure}[t]
\vspace*{-0.0cm}
\begin{center}
\hspace*{-0.0cm}\epsfysize=7truecm\epsfbox{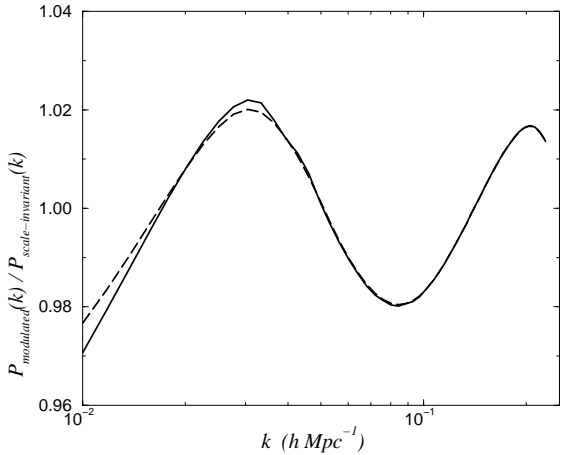}
\vspace{0truecm}
\end{center}
\vspace*{0cm}
\caption{Ratio of modulated to umodulated power spectrum, before
  (solid line) and after (dashed line) convolution with the mock 
SDSS window function.}
\label{fig:transplanckpm}
\end{figure}

With these optimistic assumptions, the effect of the window function 
is less dramatic than in the case of the 2dFGRS.  In figure
\ref{fig:transplanckpm} we show the ratio of the power spectrum 
with the standard, scale-invariant primordial $P_0(k)$ to a modulated
one with $\epsilon=0.03$, $\xi=0.023$, before (solid line) and after 
(dashed line) convolution with our mock window function.  As can be 
seen, the main effect is some smoothing of large-scale power, but the 
modulation is still clearly visible.  

\subsection{The CMB window function}

CMB data do not directly measure the underlying primordial power
spectrum of fluctuations. Rather, they measure the spectrum folded
with a transfer function in the following sense
\begin{equation}
\label{eq:cl}
C_l = \int \frac{dk}{k} P(k) \Delta_l^2 (k),
\end{equation}
where $\Delta_l^2 (k)$ is the transfer function taken at the present,
$\tau = \tau_0$, where $\tau$ is conformal time.
Following the line-of-sight approach pioneered by Seljak and Zaldariagga
\cite{CMBFAST}
this transfer function can be written as
\begin{equation}
\Delta(k) = \int_0^{\tau_0} d\tau S(k,\tau) j_l [k (\tau-\tau_0)],
\end{equation}
where $S$ is a source function, calculated from the Boltzmann equation,
and $j_l(x)$ is a spherical Bessel function.
However,
in order to get a very rough idea about the effective window function
$w_l(k) = \Delta_l^2(k)/k$ of CMB we approximate $S$ with a constant
to obtain

\begin{equation}
\label{eq:window}
w_l(k) \propto \cases{\left(\frac{k \tau_0}{l}\right)^l \sim 0 & for $\quad  k \tau_0 \lesssim l$ \cr 
\frac{1}{k^3} & for $\quad  k \tau_0 \gtrsim l$} \\
\end{equation}

This simple equation shows several things: (a) A feature at some specific
wavenumber 
$k = k_*$ has the greatest impact on the CMB spectrum 
at $l_* \simeq k_* \tau_0$.
For a flat, matter dominated universe, $\tau_0 = H_0/2$, yielding
$l_* \simeq 2 k_*/H_0$. (b) The CMB windows function is quite
broad, and narrow features in $P(k)$ are accordingly difficult to detect.

\begin{figure}[t]
\begin{center}
\epsfysize=11truecm\epsfbox{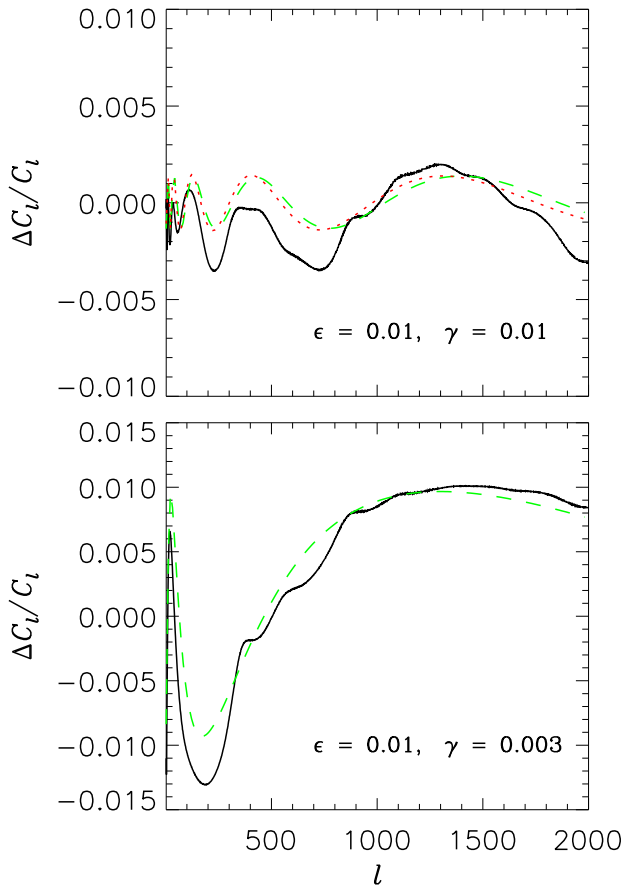}
\vspace{0truecm}
\end{center}
\vspace*{0.5cm}
\caption{Ratio of modulated to umodulated power spectrum, both the 
exact result (solid line) and the approximation Eq. (16) (dashed line).}
\label{fig:approxcl}
\end{figure}

Starting from the power spectrum given in Eq.~(\ref{eq:transplanckpk1})
with $P(k) = P_0(k)(1+\Delta P(k))$ and
\begin{equation}
\Delta P(k) = \xi \left(\frac{k}{k_0}\right)^{-\epsilon} \sin
\left[ \frac{2}{\xi} \left( \frac{k}{k_0}\right)^{\epsilon}\right]
\end{equation}
we can write the change in the CMB power spectrum, $\Delta C_l$ as
\begin{equation}
\frac{\Delta C_l}{C_l} = \frac{\int_{l/\tau_0}^{\infty} 
\frac{\Delta P(k)}{k^3} dk}{\int_{l/\tau_0}^{\infty} 
\frac{P_0(k)}{k^3} dk}
\end{equation}
Since $k_0 \simeq l/\tau_0$ this can be recast in the relatively simple form
\begin{equation}
\frac{\Delta C_l}{C_l} = \frac{1}{2 l^2} \int_l^\infty \xi q^{-\epsilon-3} \sin
\left[\frac{2}{\xi} q^{\epsilon}\right] dq.
\end{equation}
In the limit where $|\frac{2 l^{\epsilon} \epsilon}{\xi (2+\epsilon)}| \gg 1$
one can use the approximation
\begin{equation}
\frac{\Delta C_l}{C_l} = \frac{\xi^2}{\epsilon} l^{-2 \epsilon} \cos
\left[\frac{2 l^\epsilon}{\xi}\right]
\end{equation}
In figure \ref{fig:approxcl} we show the quantity $\Delta C_l/C_l$ for the same
two sets of parameters as in Fig.~2 of Ref.~\cite{danielsson2},
i.e.\ $(\epsilon = 0.01, \gamma = 0.01)$ and 
$(\epsilon = 0.01, \gamma = 0.003)$.

\subsection{Results}

\begin{figure}[t]
\vspace*{-0.0cm}
\begin{center}
\epsfysize=7truecm\epsfbox{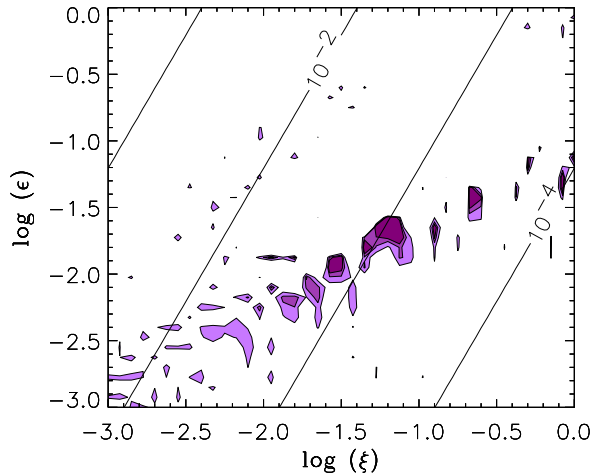}
\end{center}
\caption{68\% and 95\% confidence exclusion plot of the parameters
$\epsilon$ and $\xi$ for the simulated Planck data only.  The straight 
lines are isocontours for the parameter $\gamma$. } 
\label{fig:simcontours}
\end{figure}

\begin{figure}[t]
\vspace*{-0.0cm}
\begin{center}
\epsfysize=7truecm\epsfbox{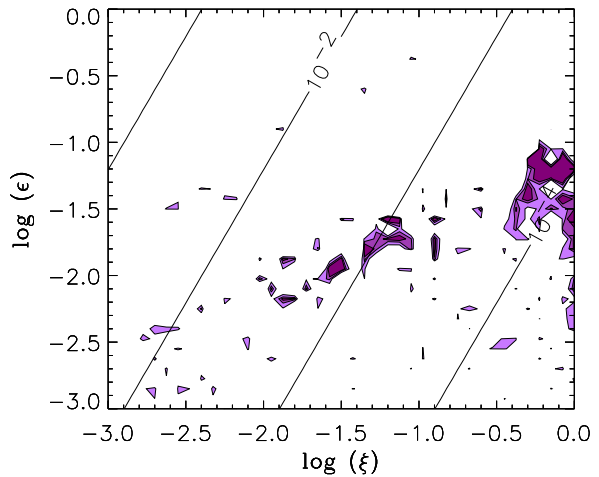}
\end{center}
\caption{68\% and 95\% confidence exclusion plot of the parameters
$\epsilon$ and $\xi$ for the simulated Planck data combined with the 
simulated SDSS data.  The straight lines are isocontours for the 
parameter $\gamma$.} 
\label{fig:simcontours2}
\end{figure}

In figures \ref{fig:simcontours} and \ref{fig:simcontours2} we show the
result of a likelihood analysis for simulated Planck and SDSS data. 
The underlying model used was a standard concordance $\Lambda$CDM model
($\Omega_m = 0.3$, $\Omega_\Lambda=0.7$, $\Omega_b h^2 = 0.024$,
$n_s=1$, and $H_0 = 70 \,\, {\rm km} \, 
{\rm s}^{-1} \, {\rm Mpc}^{-1}$),
with the addition of a modulated spectrum with parameters, 
$\epsilon=\xi=0.01$.
From the analysis of the simulated Planck data set we find that the
best fit model for $\epsilon=\xi=0.01$ has $\chi^2 = 2063.1$ for
$\nu=1992$ degrees of freedom. This is roughly within the expected 1
$\sigma$ interval of $\chi^2 = \nu \pm \sqrt{2\nu} = 1992\pm 63$.
However, the actual best fit point of the analysis was at
$\epsilon = 0.022, \xi = 0.053$ which has $\chi^2 = 2046.5$, seemingly 
ruling out the correct model at more than 99.9\% confidence.

The problem with this analysis is that both models are decent fits
within the expectations, but that the exact value of the likelihood
function is extremely sensitive to $\epsilon$ and $\xi$. This is partly
due to the fact that the phase of the spectrum oscillation at a given
$k$-value is strongly dependent on $\epsilon$ and $\xi$. It should also
be noted that any grid-based likelihood calculation is likely to fail
when faced with this type of likelihood function, and that only 
stochastic algorithms like Markov Chain Monte Carlo (MCMC) 
\cite{mcmc} or simulated
annealing \cite{Hannestad:wx} are likely to work.

\begin{figure}[t]
\vspace*{-0.0cm}
\begin{center}
\epsfysize=7truecm\epsfbox{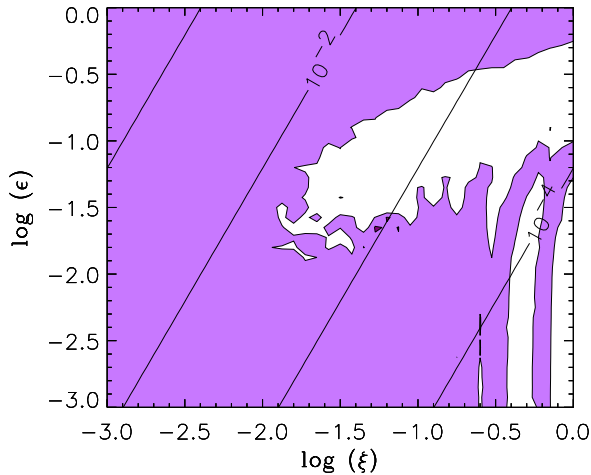}
\end{center}
\caption{68\% and 95\% confidence exclusion plot of the parameters
$\epsilon$ and $\xi$ when the $1\sigma$ region is taken to be 
$1992\pm63$, the expected $1\sigma$ interval for the $\chi^2$ distribution.
The straight lines are isocontours for the parameter $\gamma$. } 
\label{fig:simcontours3}
\end{figure}

If, instead
of using the standard prescription of assigning likelihood according to
$\Delta \chi^2$ compared with the global best fit, one uses a robust
estimation such as goodness-of-fit the plot becomes very different.
Figure \ref{fig:simcontours3} shows the contours when taking
the 1$\sigma$ region to be $1992\pm 63$, i.e.\ the expected 
1$\sigma$ interval for the $\chi^2$ distribution. In this 
case the data is unable to discriminate the underlying model from one
without any oscillation, but the huge number of non-connected regions
in the likelihood fit disappear.

This shows very clearly that the likelihood estimator is in principle
very powerful, but that it is not robust. Using a more robust
estimator significantly reduces the sensitivity, but on the other hand
also eliminates the possibility of unphysical biasing of the results.

From figure \ref{fig:simcontours3} it can also be seen that the exclusion 
region has the same topology as those using the present WMAP data
(figures 1 and 2),
and for the same reasons.

\begin{figure}[t]
\vspace*{-0.0cm}
\begin{center}
\epsfysize=11truecm\epsfbox{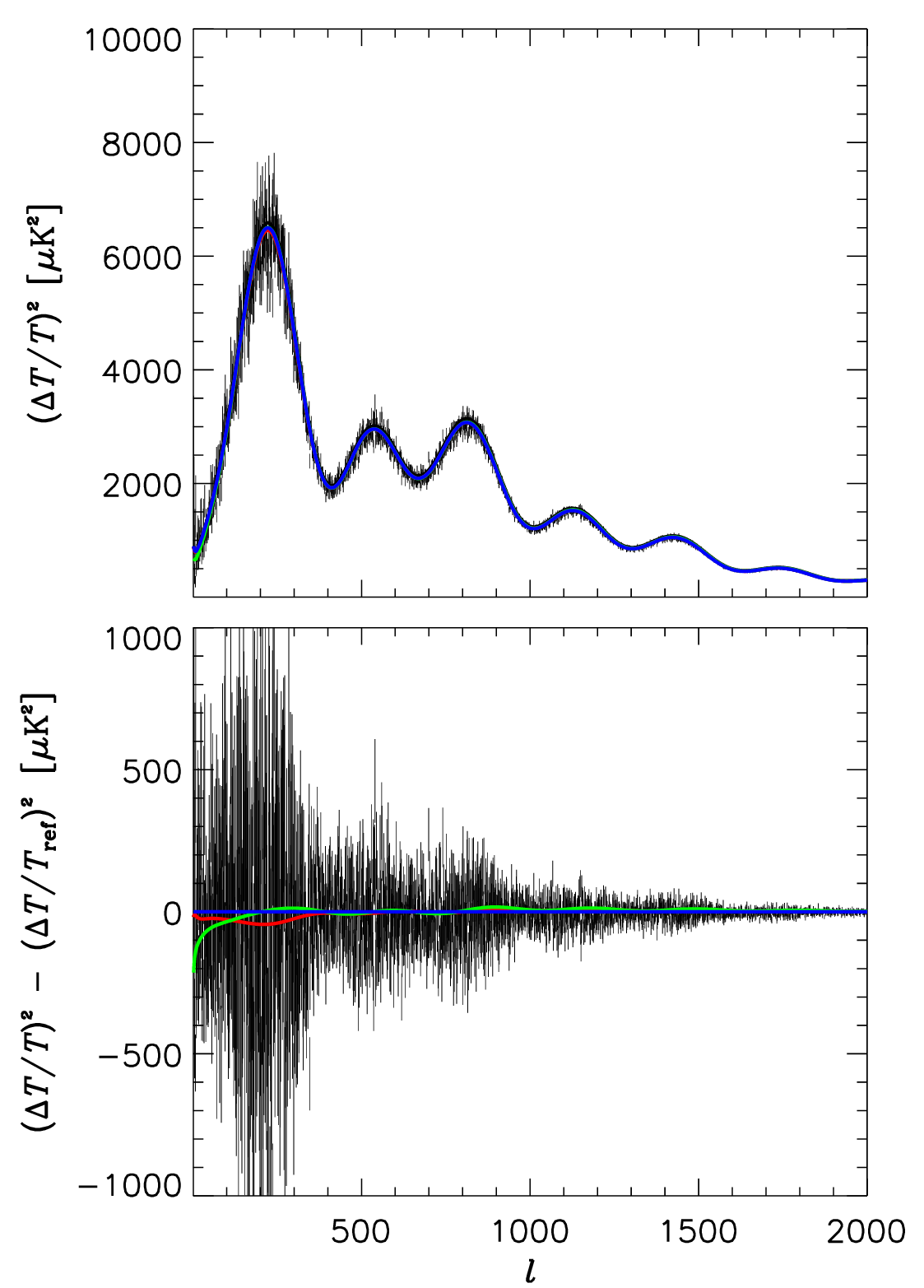}
\vspace{0.5truecm}
\end{center}
\caption{The CMB power spectra for the best fits to the simulated data 
based on the reference model $\epsilon=\xi=0.01$ (blue line).  
The green line has $\epsilon=10^{-1.2}$, 
$\xi=10^{-0.3}$, while the red line has $\epsilon=10^{-2.1}$, $\xi=10^{-1.95}$. 
The bottom panel shows the residuals with respect to the reference model} 
\label{fig:cldeg}
\end{figure}

\begin{figure}[t]
\vspace*{-0.0cm}
\begin{center}
\epsfysize=11truecm\epsfbox{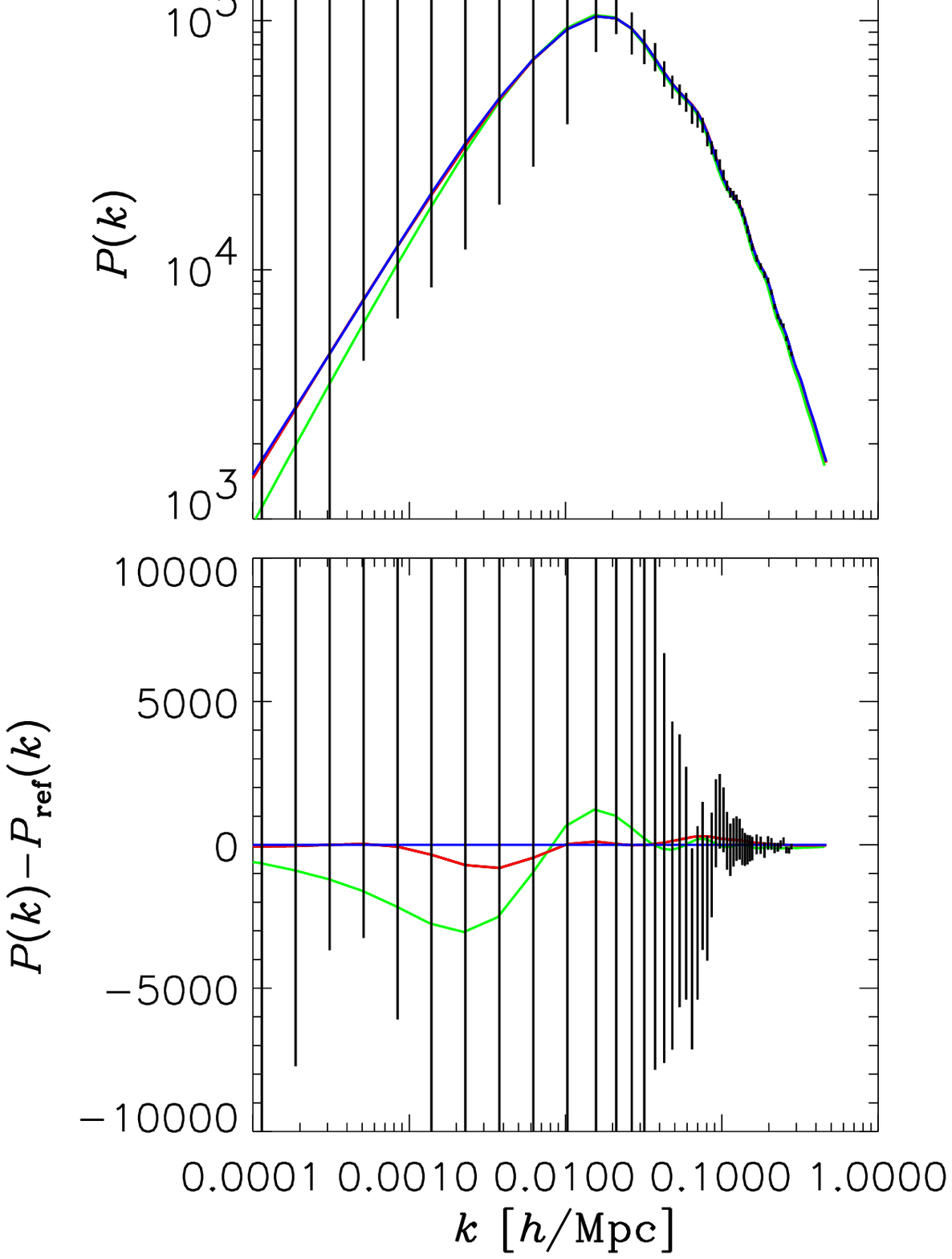}
\vspace{0.5truecm}
\end{center}
\caption{The galaxy power spectra for the best fits to the simulated data based 
on the reference model $\epsilon=\xi=0.01$ (blue line).  
The green line has $\epsilon=10^{-1.2}$, $\xi=10^{-0.3}$, while the red line 
has $\epsilon=10^{-2.1}$, $\xi=10^{-1.95}$.  The bottom panel shows the residuals 
with respect to the reference model. } 
\label{fig:pkdeg}
\end{figure}

The analysis is almost equivalent when including the mock SDSS data.
It should, however, be noted that the likelihood contours in
figures \ref{fig:simcontours} and \ref{fig:simcontours2} are very
different, showing again the sensitivity of $\Delta \chi^2$ to
$\epsilon$ and $\xi$.  Figures \ref{fig:cldeg} and \ref{fig:pkdeg} 
illustrate the degeneracy between models with widely different 
$\epsilon$ and $\xi$.

\section{Discussion}

We have performed a detailed analysis of how a logarithmic oscillation
of the primordial fluctuation spectrum influences CMB and LSS
observations. The motivation for this study was that trans-Planckian
effects could lead to exactly such a modulation of the spectrum,
but in fact our analysis of present and simulated future data sets
is more general.

The analysis of present data shows that there is no evidence at a significant
level for any oscillating behaviour of the power spectrum, but that only
a very narrow region of parameter space is currently excluded. This region
is not close to the expected values of the parameters $\epsilon$ and $\gamma$
for plausible models of Planck scale physics, such as the Horava-Witten
model.

In order to study the possibility of detecting modulations with future
data sets, such as those from the Planck satellite or the SDSS survey,
we have constructed mock data sets and performed likelihood analyses
on them. This analysis showed that the likelihood function is extremely
sensitive to the parameters $\epsilon$ and $\xi$, and that the
likelihood function exhibits a large number of distinct local minima.
This makes any analysis based on a $\Delta \chi^2$ approach very
difficult and any result is likely to be biased by unphysical effects.
On the other hand, more robust methods, such as relying on a calculation
of $\chi^2$ alone without relying on $\Delta \chi^2$ as compared to the
global best fit, are much less sensitive. We have shown for one
specific case, where parameters were chosen to be close to optimistic
estimates of the Horava-Witten model, that using a standard likelihood
analysis yields a biased parameter estimate, whereas a robust method is
unable to distinguish the underlying model from one with no oscillations.

Our conclusion is that CMB and LSS data are in principle very sensitive
to modulations in the underlying primordial power spectrum, but that
in practise it is likely to be extremely difficult to make a positive
detection of the small amplitude oscillations predicted, even with
future high precision data.

\section*{Acknowledgments}

We acknowledge use of the publicly available CMBFAST package 
written by Uros Seljak and Matias Zaldarriaga \cite{CMBFAST}.
We thank Lars Bergstr{\"o}m and Ulf Danielsson for useful comments.

\pagebreak

\newcommand\AJ[3]{~Astron. J.{\bf ~#1}, #2~(#3)}
\newcommand\APJ[3]{~Astrophys. J.{\bf ~#1}, #2~ (#3)}
\newcommand\apjl[3]{~Astrophys. J. Lett. {\bf ~#1}, L#2~(#3)}
\newcommand\ass[3]{~Astrophys. Space Sci.{\bf ~#1}, #2~(#3)}
\newcommand\cqg[3]{~Class. Quant. Grav.{\bf ~#1}, #2~(#3)}
\newcommand\mnras[3]{~Mon. Not. R. Astron. Soc.{\bf ~#1}, #2~(#3)}
\newcommand\mpla[3]{~Mod. Phys. Lett. A{\bf ~#1}, #2~(#3)}
\newcommand\npb[3]{~Nucl. Phys. B{\bf ~#1}, #2~(#3)}
\newcommand\plb[3]{~Phys. Lett. B{\bf ~#1}, #2~(#3)}
\newcommand\pr[3]{~Phys. Rev.{\bf ~#1}, #2~(#3)}
\newcommand\prog[3]{~Prog. Theor. Phys.{\bf ~#1}, #2~(#3)}

\end{document}